\documentclass[12pt]{article}
\usepackage{amsmath}
\def\a{\alpha}
\def\b{\beta}

\def\d{\delta}

\def\g{\gamma}
\def\p{\psi}

\def\be{\begin{equation}}
\def\ee{\end{equation}}
\def\arr{\begin{array}{rll}}
\def\ea{\end{array}}
\def\bea{\begin{eqnarray}}
\def\eea{\end{eqnarray}}
\def\N2{$N{=}2$}

\def\>{\rangle}
\def\<{\langle}
\def\+{\dagger}
\def\={\ =\ }

\begin{document}
%\large
\renewcommand{\thefootnote}{\fnsymbol{footnote}}
\begin{titlepage}
\setcounter{page}{0}
\begin{flushright}
LMP-TPU--3/09  \\
\end{flushright}
\vskip 1cm
\begin{center}
{\LARGE\bf Remark on quantum mechanics with  }\\
\vskip 0.5cm
{\LARGE\bf $\mathcal{N}=2$ Schr\"odinger supersymmetry }\\
\vskip 2cm
$
\textrm{\Large Anton Galajinsky and Ivan Masterov\ }
$
\vskip 0.7cm
{\it
Laboratory of Mathematical Physics, Tomsk Polytechnic University, \\
634050 Tomsk, Lenin Ave. 30, Russian Federation} \\
{Emails: galajin@mph.phtd.tpu.ru, masterov@mph.phtd.tpu.ru}

\end{center}
\vskip 1cm
\begin{abstract} \noindent

\end{abstract}
A unitary transformation which relates
a many--body quantum mechanics with
$\mathcal{N}=2$ Schr\"odinger supersymmetry
to a set of decoupled superparticles is proposed.  The simplification in dynamics is achieved at a price of a
nonlocal realization of the full $\mathcal{N}=2$ Schr\"odinger superalgebra in a Hilbert space.
The transformation is shown to be universal and
applicable to conformal many--body models confined in a harmonic trap.


\vskip 0.5cm


\end{titlepage}

\renewcommand{\thefootnote}{\arabic{footnote}}
\setcounter{footnote}0

\noindent
{\bf 1. Introduction}\\
\noindent

Quantum mechanics of particles interacting via a conformal potential in arbitrary dimension and confined in a harmonic
trap has been extensively studied over the last three decades (see \cite{cm}--\cite{gala1} and references therein).
On the one hand, this line of research is motivated by the desire to construct new exactly solvable many--body models
in higher dimensions and explore novel correlations. On the other hand, some of these systems are of real
physical interest \cite{njlq,njlq1}.
More recently, several proposals for a nonrelativistic version of the AdS/CFT
correspondence \cite{son1}--\cite{ad} \footnote{By now there is an extensive literature on the subject. For a more complete list
of references see e.g. \cite{sch,bgop}.} stimulated a
renewed interest in the nonrelativistic conformal symmetry. It is natural to expect that many--body models with
conformal symmetry will provide new insight into the correspondence.

Galilei algebra can be extended by conformal generators in two different ways. The first option is to add the generators of
dilatations and special conformal transformations which form $so(1,2)$ subalgebra together with the generator of time translations.
The resulting algebra is known as the Schr\"odinger algebra \cite{nied}--\cite{jackiw}. Another option is to consider a
contraction of the relativistic conformal algebra $so(d+1,2)$ (see e.g. \cite{bgop} and references therein).
This yields a larger algebra which goes under the name conformal Galilei algebra. Throughout this work we discuss
the Schr\"odinger algebra.

$so(1,2)$ is the conformal algebra in one dimension.
It was realized long ago \cite{gar} that
the conformal subalgebra of the Schr\"odinger algebra plays a distinguished role.
It can be viewed as a spectrum generating algebra which allows one to construct
some excited states in explicit form \cite{gar},\cite{gh2}--\cite{msbg}.
It is also natural to
expect that quantum mechanics invariant under the Schr\"odinger
group may share some interesting features with one--dimensional
conformal systems, the most  interesting example being the
Calogero model \cite{calo1}.

In \cite{pol,glp} a unitary transformation
relating the conformal Calogero model
to a set of decoupled particles was constructed. It generalized a mapping
of the Calogero model with the oscillator potential to a set of decoupled
oscillators previously found in \cite{gur}. In \cite{gala1} the decoupling transformation
was extended to conformal many--body systems in arbitrary dimension.
In essence the decoupling transformation is a specific automorphism of the Schr\"odinger
algebra generated by the conformal subalgebra $so(1,2)$.
It should be mentioned that the simplification
in dynamics is achieved at the price of a nonlocal realization of the
full Schr\"odinger algebra in a Hilbert space \cite{gala1}. An interesting  geometric
interpretation of the decoupling transformation as the inversion of the Klein
model of the Lobachevsky plane was proposed in \cite{arm,bn}.

The purpose of this work is to generalize the results obtained  in \cite{gala1}
in two directions. After exposing some technical details on the decoupling transformation in the next section,
in section 3 we construct a similarity transformation which relates a generic conformal many--body system
confined in a harmonic trap to a set of decoupled oscillators. This allows one to draw a general conclusion
on the spectrum and quantum integrability. Then in section 4 we construct a decoupling unitary transformation for a generic quantum many--body
model invariant under $\mathcal{N}=2$ Schr\"odinger supergroup\footnote{Recent results of \cite{wyl}--\cite{klp}
on $\mathcal{N}=4$ superconformal many--body models in one dimension indicate that the $\mathcal{N}=2$
superextension is likely to be the maximal one compatible with translation invariance.}.

A few words about notation. For simplicity we focus on three spatial dimensions.
A generalization to higher dimensions is straightforward.
Greek letters are reserved to label $N$ identical particles $\a=1,\dots,N$.
Spacial indices will be almost always omitted with the obvious notation $x_\a^i=\mathbf{x}_\a$, $x^i_\a x^i_\a=\mathbf{x}_\a^2$, $\epsilon_{ijk} x^j_\a p^k_\a=\mathbf{x}_\a \times \mathbf{p}_\a$, where $\epsilon_{ijk}$ is the totaly antisymmetric tensor with
$\epsilon_{123}=1$. Throughout the work the summation over repeated indices is understood.

\vspace{0.5cm}

\noindent
{\bf 2. Decoupling transformation on conformal mechanics}\\
\noindent

Schr\"odinger algebra involves the generator of time translations $H$, the generators of space translations
$\mathbf{P}$ and space rotations $\mathbf{J}$, the generator of Galilei boosts $\mathbf{K}$ as well as
those of dilatations $D$ and special conformal transformations $C$ \cite{nied}--\cite{jackiw} (see also a related work \cite{hor1}).
The non-vanishing commutation relations in the algebra read
\bea\label{alg}
&&
[J_i,J_j]=i\hbar \epsilon_{ijk} J_k, \qquad
[J_i,P_j]=i\hbar \epsilon_{ijk} P_k, \qquad [J_i,K_j]=i\hbar \epsilon_{ijk} K_k, \qquad [K_i,P_j]=i\hbar \d_{ij} M,
\nonumber\\[2pt]
&&
[H,K_i]=-i\hbar P_i\ ,\qquad [D,K_i]=\frac{i\hbar}{2}  K_i, \qquad \quad
[C,P_i]=i \hbar K_i,\qquad \qquad [D,P_i]=-\frac{i\hbar }{2} P_i,
\nonumber\\[2pt]
&&
[H,D]=i\hbar H, ~~\quad \quad\quad [H,C]=2i\hbar D, \qquad \quad ~[D,C]=i \hbar C,
\eea
where $M$ is the central charge.
The operators $H$, $D$, and $C$ obey the commutation relations of $so(1,2)$ which is the conformal algebra in one dimension.
Other generators visualize higher dimensions.

In order to construct a quantum mechanical representation of this algebra, one introduces the position operators $\mathbf{x}_\a$
and the momentum operators $\mathbf{p}_\a$ of $N$ identical particles of mass $m$ interacting via a potential $V(\mathbf{x_1},\dots,\mathbf{x}_N)$ and constructs the
following quantities\footnote{We work in the
standard coordinate representation in which $p_\a^i=-i \hbar \frac{\partial}{\partial x_\a^i}$ and
$[x_\a^i, p_\b^j]=i\hbar \d^{ij} \d_{\a\b}$. The time dependent pieces are kept explicit so as to have
a direct link to a classical theory.}
\bea\label{gen}
&&
H=\frac{1}{2 m} \mathbf{p}_\a \mathbf{p}_\a+V(x), \qquad ~
\mathbf{P}=\sum_{\a=1}^N \mathbf{p}_\a, ~\qquad
\mathbf{J}=\mathbf{x}_\a \times \mathbf{p}_\a,~\qquad \mathbf{K}=\sum_{\a=1}^N m \mathbf{x}_\a\ -t \mathbf{P},
\nonumber\\[2pt]
&&
D=tH-\frac{1}{4}(\mathbf{x}_\a \mathbf{p}_\a+\mathbf{p}_\a \mathbf{x}_\a), \qquad
C=-t^2 H+2t D+\frac{1}{2} m \mathbf{x}_\a^2, \qquad M=mN.
\eea
The commutation relations (\ref{alg}) constrain the potential $V(x)$ to obey the system of partial differential equations
\bea\label{str}
\left(x_\a^i \frac{\partial}{\partial x_\a^j}-x_\a^j
\frac{\partial}{\partial x_\a^i}\right)V(x)=0, \qquad
\sum_{\a=1}^N \frac{\partial V(x)}{\partial x_\a^i}=0,\qquad
x_\a^i \frac{\partial V(x)}{\partial x_\a^i}+2 V(x)=0.
\eea
Any solution to (\ref{str}) defines a multi--particle quantum mechanics with the Schr\"odinger symmetry.

The simplest solution $V(x)=0$ corresponds to decoupled
particles which are governed by the free Hamiltonian.
In what follows we shall use the notation
\be\label{nota}
H_0=\frac{1}{2 m} \mathbf{p}_\a \mathbf{p}_\a, \qquad C_0=\frac{1}{2} m \mathbf{x}_\a^2, \qquad
D_0=-\frac{1}{4}(\mathbf{x}_\a \mathbf{p}_\a+\mathbf{p}_\a \mathbf{x}_\a).
\ee
Setting a zero value
for the parameter $t$ in (\ref{gen})
one does not spoil the algebra. Thus,
both $(H,D_0,C_0)$ and $(H_0,D_0,C_0)$ obey the commutation relations of $so(1,2)$ provided
the rightmost equation in (\ref{str}) holds.
This simple fact will prove to be important in our subsequent consideration.

Consider an automorphism of the Schr\"odinger algebra
\be\label{ser}
T \quad \rightarrow \quad T'=e^{\frac{i}{\hbar}A}~ T~ e^{-\frac{i}{\hbar}A}=T+\sum_{n=1}^\infty\frac{1}{n!}{\left(\frac{i}{\hbar}\right)}^n
\underbrace{[A,[A, \dots [A,T] \dots]}_{n~\rm times} \ ,
\ee
generated by a specific linear combination of the operators $H$, $D_0$, $C_0$
\be\label{a}
A\=\a H + \frac{1}{\a} C_0 -2D_0.
\ee
Here $\a$ is an arbitrary parameter of the dimension of length. The number
coefficients in (\ref{a}) are adjusted so as to terminate the infinite
series in the Baker--Hausdorff formula  (\ref{ser}) at a final step. To be more precise, one takes $A$ to be a
linear combination  of $H$, $D_0$, $C_0$ with arbitrary coefficients and then demands the series to terminate at most at the third step.
This fixes one coefficient in terms of the others. Then one requires $H$ to disappear from $H'$ which
determines one more coefficient and yields (\ref{a}). For the generators of the Schr\"odinger algebra one finds
\bea\label{inter}
&&
\mathbf{J}'=\mathbf{J}, \qquad
\mathbf{P}'=-\frac{1}{\a} (\mathbf{K}+t \mathbf{P}), \qquad \mathbf{K}'=\left(2+\frac{t}{\a}\right) \mathbf{K}+
\a{\left(1+\frac{t}{\a}\right)}^2 \mathbf{P},
\nonumber\\[2pt]
&&
H'=\frac{1}{\a^2} C_0, \qquad C'=\a^2 (H_0+V(x))-2\a\left(2+\frac{t}{\a} \right)D_0+{\left(2+\frac{t}{\a} \right)}^2 C_0,
\nonumber\\[2pt]
&&
D'=-D_0+\frac{1}{\a}\left(2+\frac{t}{\a} \right) C_0.
\eea

That the interacting Hamiltonian is mapped to $C_0$ prompts one to
try a similar transformation involving $(H_0,D_0,C_0)$
\be\label{sec}
T' \quad \rightarrow \quad T^{''}=e^{\frac{i}{\hbar}B}~ T'~ e^{-\frac{i}{\hbar}B}, \qquad B=-\a H_0-\frac{1}{\a} C_0+2 D_0.
\ee
Such a choice of $B$ again terminates the infinite series at most at the third step and yields the result
\bea
&&
H''=H_0\, \qquad D''=tH_0+D_0\ , \qquad
C''=-t^2 H_0+2t D''+C_0+\a^2 \left( e^{\frac{i}{\hbar} B}~V(x)~ e^{-\frac{i}{\hbar} B}\right),
\nonumber\\[2pt]
&&
\mathbf{J}''=\mathbf{J}, \qquad \mathbf{P}''=\mathbf{P}, \qquad \mathbf{K}''=\mathbf{K}.
\eea
To summarize, with the use of the unitary operator $e^{\frac{i}{\hbar} B} e^{\frac{i}{\hbar} A}$ one can map
the Hamiltonian of a generic many--body system with the Schr\"odinger symmetry to that of $N$ decoupled particles.
Notice, however, that the interaction potential $V(x)$
reappears in the generator of special conformal transformations as a nonlocal contribution.
Thus the simplification in dynamics is achieved at
the price of a nonlocal realization of the full Schr\"odinger algebra in a
Hilbert space of a resulting quantum mechanical model.

Concluding this section let us demonstrate that $e^{\frac{i}{\hbar} B} e^{\frac{i}{\hbar} A}$ does not depend on the dimensionful
parameter $\a$. This can be verified by computing the derivative of the operator with respect to $\a$ and showing that it is zero.
Taking into account the explicit form of
$B$ and the commutation relations among $(H_0,C_0,D_0)$
one finds the relations
\be
[\frac{d B}{d \a}, B^n]= \frac{2i\hbar}{\a} n B^n \quad \Rightarrow \quad \frac{d B^n}{d\a}=
n\left(  \frac{d B}{d\a}\right) B^{n-1} -i n(n-1) \frac{\hbar}{\a}B^{n-1}.
\ee
With these at hand, it is straightforward to derive the formula
\be\label{b1}
\frac{d}{d\a} \left( e^{\frac{i}{\hbar}B}\right) = \frac{i}{\hbar}\left( \frac{d B}{d \a} +\frac{1}{\a} B \right)e^{\frac{i}{\hbar}B}=\frac{2i}{\a\hbar} \left(D_0-\a H_0 \right) e^{\frac{i}{\hbar}B}=\frac{2i}{\a^2\hbar} e^{\frac{i}{\hbar}B} \left(C_0-\a D_0 \right).
\ee
The operator $A$ can be treated in a similar fashion
\be\label{a1}
[\frac{d A}{d \a}, A^n]=- \frac{2i\hbar}{\a} n A^n \quad \Rightarrow \quad
\frac{d}{d\a} \left( e^{\frac{i}{\hbar}A}\right) = \frac{i}{\hbar}\left( \frac{d A}{d \a} -\frac{1}{\a} A \right)e^{\frac{i}{\hbar}A}=
-\frac{2i}{\a^2 \hbar}
\left(C_0-\a D_0\right) e^{\frac{i}{\hbar}A}.
\ee
Along with (\ref{b1}) this leads to the desired result.

\vspace{0.5cm}

\noindent
{\bf 3. Conformal mechanics in a harmonic trap}\\

From a physical point of view a more realistic model arises when one considers a many--body
conformal mechanics in a confining harmonic potential
\be\label{harm}
\mathcal{H}=H_0+V(x)+\frac {m \omega^2}{2N} \sum_{\a<\b}^N {(\mathbf{x_\a}-\mathbf{x_\b})}^2,
\ee
where $\omega$ is the frequency. Notice that the last term fails
to obey the rightmost equation in (\ref{str}). Let us demonstrate
that the method outlined in the previous section can be used to construct a similarity
transformation relating (\ref{harm}) to $(N-1)$ decoupled oscillators.

The idea is to go over to new coordinates in which the harmonic potential can be identified with the operator $C_0$
from the previous section.
This makes feasible an implementation of the algebraic construction considered above in a more complicated setting.
It turns out that the Jacobi coordinates
\bea
&&
\mathbf{y}_a=\frac{1}{\sqrt{a(a+1)}} \left(\sum_{\b=1}^a \mathbf{x}_\b-a \mathbf{x}_{a+1}\right), \qquad
{\mathbf{w}}_a=\frac{1}{\sqrt{a(a+1)}} \left(\sum_{\b=1}^a \mathbf{p}_\b-a \mathbf{p}_{a+1}\right),
\nonumber\\[2pt]
&&
\mathbf{X}=\frac 1N \sum_{\a=1}^N \mathbf{x}_\a, \qquad \qquad \qquad \qquad \qquad \quad ~  \mathbf{P}=\sum_{\a=1}^N \mathbf{p}_\a,
\eea
where $a=1,\dots,N-1$ are sufficient to achieve the aim. Notice that the center of mass coordinates $(\mathbf{X},\mathbf{P})$ and
the relative motion coordinates $(\mathbf{y}_a,\mathbf{w}_a)$ obey canonical commutation relations.
Being rewritten in the new coordinates, the Hamiltonian reads
\be
\mathcal{H}=\frac {1}{2M} \mathbf{P}^2+\frac{1}{2m} \mathbf{w}_a^2+V(\mathbf{y})+\frac {m \omega^2 }{2} \mathbf{y}_a^2,
\ee
where $M=mN$ is the mass of the system.

Because the center of mass motion is separated, one can focus on the relative coordinates and introduce the operators
similar to those above
\bea
&&
D_0^{(rel)}=-\frac{1}{4} (\mathbf{y}_a \mathbf{w}_a+\mathbf{w}_a \mathbf{y}_a), \quad \qquad
C_0^{(rel)}=\frac{m}{2} \mathbf{y}_a^2=\frac {m}{2N} \sum_{\a<\b}^N {(\mathbf{x_\a}-\mathbf{x_\b})}^2,
\nonumber\\[2pt]
&&
H^{(rel)}_0=\frac{1}{2m} \mathbf{w}_a^2, \quad \qquad H=H^{(rel)}_0+V(\mathbf{y}), \quad \qquad H^{(rel)}=H+\omega^2 C_0^{(rel)}.
\eea
It is worth mentioning that both $(H^{(rel)}_0,D^{(rel)}_0,C^{(rel)}_0)$ and $(H,D^{(rel)}_0,C^{(rel)}_0)$ obey the commutation relations of $so(1,2)$\footnote{In the original coordinates for a conformal potential $V(x)$ one has $[V(x),D_0]=i\hbar V(x)$. Because the potential is
translation invariant, i.e. $[V(x),\mathbf{P}]=0$, one gets a similar relation $[V(\mathbf{y}),D_0^{(rel)}]=i\hbar V(\mathbf{y}) $
after passing to the Jacobi coordinates.}.
Then one can consider a similarity transformation of the full Hamiltonian which affects only the relative motion coordinates
\be\label{ps}
\mathcal{H}''=e^{\frac{i}{\hbar} B} e^{\frac{i}{\hbar} A}  \mathcal{H} e^{-\frac{i}{\hbar} A} e^{-\frac{i}{\hbar} B}=
\frac {1}{2M} \mathbf{P}^2+e^{\frac{i}{\hbar} B} e^{\frac{i}{\hbar} A} H^{(rel)} e^{-\frac{i}{\hbar} A} e^{-\frac{i}{\hbar} B}.
\ee
Here $A$ is supposed to be a linear combination of $(H,D^{(rel)}_0,C^{(rel)}_0)$ with arbitrary coefficients, while $B$ is constructed
from $(H^{(rel)}_0,D^{(rel)}_0,C^{(rel)}_0)$. Like in the preceding section, the arbitrary coefficients in $A$ and $B$ are fixed from the
requirement that
the series in (\ref{ps}) terminates at the third step and that the conformal potential disappears from the transformed Hamiltonian. It is straightforward to verify that the operators
\be
A=\a H+\frac{1}{\a}{(1-i\a\omega)}^2 C^{(rel)}_0-2(1-i\a\omega)D^{(rel)}_0,
\ee
and
\be
B=-\a H^{(rel)}_0-\frac{1}{\a}{(1-i\a\omega)}^2 C^{(rel)}_0+2(1-i\a\omega)D^{(rel)}_0,
\ee
realize the desired map
\be
\mathcal{H}''=\frac {1}{2M} \mathbf{P}^2+\frac{1}{2m} \mathbf{w}_a^2+\frac{m \omega^2}{2} \mathbf{y}_a^2.
\ee
In contrast to purely conformal models, the similarity transformation is provided
by a non--unitary operator. This correlates well with the results obtained previously for one--dimensional systems \cite{glp,gur}.

To summarize, a conformal mechanics of $N$ identical particles confined in a harmonic trap can be mapped to a set of $N-1$ decoupled oscillators
(plus the center of mass). An immediate corollary is that the energy eigenvalues are the same in both the pictures.
Another corollary is that the original many--body model is quantum integrable. This can be argued as follows.
After the map one has a set of decoupled oscillators. Kinetic energies of individual oscillators are mutually
commuting. They also commute with the full Hamiltonian and the kinetic energy of the center of mass.
Taking the inverse map of these operators one gets $N$ independent and mutually commuting conserved operators for the original
interacting system.

\vspace{0.5cm}

\noindent
{\bf 4. $\mathcal{N}=2$ supersymmetric extension}\\

Our next objective is to generalize the analysis in section 2 to the case of many--body quantum mechanics with  $\mathcal{N}=2$
Schr\"odinger supersymmetry \footnote{For a review of non--relativistic
supersymmetry see e.g. \cite{hor}--\cite{dh1} and references therein.}.
Like in one dimension our count of supersymmetries corresponds to the number of self--adjoint supercharges.
Taking fermionic degrees of freedom $\psi^i$ to be self--adjoint ${\left(\psi^{i}\right)}^{\dagger}=\psi^i$ and forming a Clifford algebra $\{\psi^i,\psi^j \}=\delta^{ij}$, one can construct one self--adjoint supersymmetry operator
which enters $\mathcal{N}=1$  Schr\"odinger
superalgebra \cite{hor}. In this case the fermionic degrees of freedom are represented by the Pauli matrices and quantum states
carry spinor indices. If instead fermionic degrees of freedom are chosen to be complex, i.e. ${\left(\psi^{i}\right)}^{\dagger}=\bar\psi^i$, and obeying the algebra $\{\psi^i,\psi^j\}=0$, $\{{\bar\psi}^i,{\bar\psi}^j\}=0$, $\{\psi^i,{\bar\psi}^j\}=\delta^{ij}$, then one can construct two supercharges which are conjugates of each other and enter $\mathcal{N}=2$ Schr\"odinger superalgebra. In the latter case the fermionic operators act as creation--annihilation operators in an appropriate Fock space.
Assigning an extra spinor index to fermions $\psi^i \rightarrow  \psi^i_\alpha$ would lead to $\mathcal{N}=4$ models.
It should be stressed that the supersymmetry considered in this work is not conventional in the sense that
its relativistic analogue is likely to be an $\mathcal{N}=2$ variant of the space-time {\it vector} supersymmetry
introduced in \cite{bard} and recently discussed in
\cite{casal}.

Apart from the generators considered above, $\mathcal{N}=2$
Schr\"odinger superalgebra includes one extra bosonic operator $R$ which corresponds to $u(1)$ R-symmetry. Fermionic operators involve a pair of supersymmetry generators $Q$, $\bar Q$, the superconformal generators $S$, $\bar S$, which can be viewed as fermionic partners of $C$, as well as the partners of the Galilei boosts $\mathbf{L}$ and $\bar{\mathbf{L}}$. It is assumed that the odd operators are hermitian conjugates of each other
\be
{Q}^{\dagger}=\bar Q, \qquad {S}^{\dagger}=\bar S, \qquad {\mathbf{L}}^{\dagger}=\bar{\mathbf{L}}.
\ee
Along with (\ref{alg}) the non--vanishing (anti)commutation relations in the algebra include
\bea\label{n2}
&&
\{Q,\bar Q \}=2\hbar H, \qquad [Q,K_i]=-i\hbar L_i, \qquad [\bar Q,K_i]=-i\hbar \bar{L}_i, \qquad \{Q,\bar L_i \}=\hbar P_i,
\nonumber\\[2pt]
&&
\{\bar Q,L_i \}=\hbar P_i, \qquad \{L_i,\bar L_j\}=\hbar \delta_{ij} Z_1, \qquad [J_i,L_j]=i\hbar \epsilon_{ijk} L_k, \quad
[J_i,\bar L_j]=i\hbar \epsilon_{ijk} \bar L_k,
\nonumber\\[2pt]
&&
[Q,D]=\frac{i\hbar}{2} Q, \qquad ~ [Q,C]=-i\hbar S, \qquad \quad [\bar Q,D]=\frac{i\hbar}{2} \bar Q, \qquad \quad [\bar Q,C]=-i\hbar \bar S,
\nonumber\\[2pt]
&&
[H,S]=-i\hbar Q, \qquad [H,\bar S]=-i\hbar \bar Q, \qquad \quad  [S,P_i]=i\hbar L_i, \qquad \quad [\bar S,P_i]=i\hbar \bar L_i,
\nonumber\\[2pt]
&&
\{S,\bar L_i\}=\hbar K_i, \qquad \{\bar S,L_i\}=\hbar K_i, \qquad \quad ~~ [D,S]=\frac{i\hbar}{2} S, \qquad \quad  [D,\bar S]=\frac{i\hbar}{2} \bar S,
\nonumber\\[2pt]
&&
[R,Q]=\hbar Q, \qquad \quad [R,\bar Q]=-\hbar \bar Q, \quad \qquad ~~ [R,S]=\hbar S, \qquad \quad~~ [R,\bar S]=-\hbar \bar S,
\nonumber\\[2pt]
&&
\{S,\bar S\}=2\hbar C, \qquad ~ [R,L_i]=\hbar L_i, \qquad \quad \quad [R,\bar L_i]=-\hbar \bar L_i,
\nonumber\\[2pt]
&&
\{Q,\bar S \}=-2\hbar D-i\hbar R+i\hbar Z, \qquad \qquad \quad ~~  \{\bar Q,S \}=-2\hbar D+i\hbar R-i\hbar Z,
\eea
where $Z$ and $Z_1$ are the central charges.

In order to build a quantum mechanical representation of this algebra, one introduces the operators {\boldmath{$\p_\a$}}=${\psi}^i_\a$ and {\boldmath{$\bar\p_\a$}}=${\bar\psi}^i_\a$ which describe the fermionic degrees of freedom and obey
\be
\{\p^i_\a,{\bar\p}^j_\b \}=\hbar \delta^{ij} \delta_{\a\b}, \qquad \{\p^i_\a,\p^j_\b \}=0, \qquad \{{\bar\p}^i_\a,{\bar\p}^j_\b \}=0, \qquad {(\psi^i_\a)}^{\dagger}={\bar\psi}^i_\a.
\ee
With these at hand, the generators can be constructed in terms of a single prepotential $U(\mathbf{x}_1, \dots,\mathbf{x}_N)$ by analogy with conventional one--particle supersymmetric quantum mechanics
\bea\label{gener}
&&
Q=\frac{1}{\sqrt{m}}{\mbox{\boldmath{$\p$}}}_\a (\mathbf{p}_\a+i \sqrt{m} \mathbf{u}_\a), \qquad
\bar Q= \frac{1}{\sqrt{m}}{\mbox{\boldmath{$\bar\p$}}}_\a (\mathbf{p}_\a-i \sqrt{m} \mathbf{u}_\a),
\nonumber\\[2pt]
&&
S=\sqrt{m} \mathbf{x}_\a {\mbox{\boldmath{$\p$}}}_\a-t Q, \qquad \quad \quad \quad
\bar S=\sqrt{m}\mathbf{x}_\a {\mbox{\boldmath{$\bar\p$}}}_\a-t \bar Q,
\nonumber\\[2pt]
&&
\mathbf{L}=\sqrt{m} \sum_{\a=1}^N {\mbox{\boldmath{$\p$}}}_\a, \qquad \qquad \qquad \quad  \mathbf{\bar L}=\sqrt{m} \sum_{\a=1}^N {\mbox{\boldmath{$\bar\p$}}}_\a,
\nonumber\\[2pt]
&&
\mathbf{P}=\sum_{\a=1}^N \mathbf{p}_\a, \qquad \qquad \qquad \qquad \quad
\mathbf{K}=\sum_{\a=1}^N m \mathbf{x}_\a\ -t \mathbf{P},
\nonumber\\[2pt]
&&
R= <{\mbox{\boldmath{$\p$}}}_\a {\mbox{\boldmath{$\bar\p$}}}_\a>,  ~ \qquad \qquad \quad  \qquad
\mathbf{J}=\mathbf{x}_\a \times \mathbf{p}_\a-i{\mbox{\boldmath{$\p$}}}_\a \times {\mbox{\boldmath{$\bar\p$}}}_\a,
\nonumber\\[2pt]
&&
D=tH-\frac{1}{4}(\mathbf{x}_\a \mathbf{p}_\a+\mathbf{p}_\a \mathbf{x}_\a), \qquad
C=-t^2 H+2t D+\frac{1}{2} m \mathbf{x}_\a^2,
\nonumber\\[2pt]
&&
H=\frac{1}{2 m} \mathbf{p}_\a^2+\frac 12 \mathbf{u}_\a^2-\frac{1}{\sqrt{m}}<{\mbox{\boldmath{$\p$}}}_\a {\mbox{\boldmath{$\bar\p$}}}_\b> \mathbf{u}_{\a\b}=H_0+V,
\eea
where we denoted $\mathbf{u}_\a=u^i_\a=\frac{\partial U(\mathbf{x}_1, \dots,\mathbf{x}_N)}{\partial x^i_\a} $ and
$<${\boldmath{$\p_\a$}} {\boldmath{$\bar\p_\b$}}$>$ $\mathbf{u}_{\a\b}=\frac 12 (\p^i_\a \bar\p^j_\b-\bar\p^j_\b \p^i_\a) \frac{\partial^2 U(\mathbf{x}_1,\dots,\mathbf{x}_N)}{\partial x^i_\a \partial x^j_\b}$ and fixed two central charges $Z_1=M=mN$.
Notice that the generators are Weyl--ordered.

The operators (\ref{gener}) form a representation of $\mathcal{N}=2$ Schr\"odinger superalgebra provided the prepotential $U(\mathbf{x}_1,\dots,\mathbf{x}_N)$ satisfies the set of partial differential equations
\bea\label{str1}
\left(x_\a^i \frac{\partial}{\partial x_\a^j}-x_\a^j
\frac{\partial}{\partial x_\a^i}\right)U=0, \qquad
\sum_{\a=1}^N \frac{\partial U}{\partial x_\a^i}=0,\qquad
\sqrt{m} x_\a^i \frac{\partial U}{\partial x_\a^i}=Z,
\eea
where $Z$ is the central charge from (\ref{n2}). Any solution to (\ref{str1}) yields a many--body quantum mechanics
invariant under $\mathcal{N}=2$ Schr\"odinger supergroup.
Notice that in the bosonic limit one reveals the potential
\be
V_B=\frac 12 \mathbf{u}_\a^2.
\ee
That the latter is conformal is guaranteed by the rightmost equation in (\ref{str1}).
For example, a prepotential of the form
\be
U=\frac{g}{2} \sum_{\a<\b}\ln|\mathbf{r}_{\a\b}^2|,
\ee
where $\mathbf{r}_{\a\b}=\mathbf{x}_\a-\mathbf{x}_\b$ and $g$ is a coupling constant yields
\be
V_B=g^2 \left( \sum_{\a<\b} \frac{1}{\mathbf{r}_{\a\b}^2}+\sum_{\g,\g\ne\a,\g\ne\b} \sum_{\a<\b} \frac{\mathbf{r}_{\g\a} \mathbf{r}_{\g\b}}{\mathbf{r}_{\g\a}^2 \mathbf{r}_{\g\b}^2}\right).
\ee
This is the model of Calogero and Marchioro \cite{cm} in which one
disregards
the pairwise harmonic interaction and identifies the couplings which control the two--body and three--body interactions \footnote{This identification is typical for superconformal mechanics (see e.g. the discussion in \cite{gala}).}.

In order to construct a unitary transformation which maps a generic many--body quantum mechanics
with $\mathcal{N}=2$ Schr\"odinger supersymmetry
to a set of decoupled superparticles, we proceed as described in section 2.
Firstly, along with (\ref{nota}) one introduces the fermionic operators
\be
Q_0=\frac{1}{\sqrt{m}}{\mbox{\boldmath{$\p$}}}_\a \mathbf{p}_\a, \qquad {\bar Q}_0=\frac{1}{\sqrt{m}}{\mbox{\boldmath{$\bar\p$}}}_\a \mathbf{p}_\a,
\qquad S_0=\sqrt{m}\mathbf{x}_\a {\mbox{\boldmath{$\p$}}}_\a, \qquad
{\bar S}_0=\sqrt{m}\mathbf{x}_\a {\mbox{\boldmath{$\bar\p$}}}_\a.
\ee
Secondly, one considers two transformations with $A$ and $B$ as in (\ref{a}) and (\ref{sec}), respectively. $H$ entering $A$ is assumed to be the full supersymmetric Hamiltonian from (\ref{gener}). Finally, it is a matter of straightforward, although a bit tedious,
calculation to verify that a consecutive application of the $A$-- and $B$--maps yields
\bea
&&
Q''=Q_0, \qquad \quad \bar Q''=\bar Q_0, \quad \qquad H''=H_0, \qquad \quad D''=tH_0+D_0, \quad \qquad R''=R,
\nonumber\\[2pt]
&&
\mathbf{J}''=\mathbf{J}, \qquad \quad  \quad \mathbf{P}''=\mathbf{P}, \quad \qquad ~ \mathbf{K}''=\mathbf{K}, ~ \qquad \quad  \mathbf{L}''=\mathbf{L}, \qquad \qquad \quad  \quad ~ \mathbf{\bar L}''=\mathbf{\bar L},
\nonumber\\[2pt]
&&
S''=S_0-t Q_0-\frac{i \a}{\hbar} e^{\frac{i}{\hbar}B}~[S_0,V]~ e^{-\frac{i}{\hbar}B}, \qquad ~
\bar S''=\bar S_0-t \bar Q_0-\frac{i \a}{\hbar} e^{\frac{i}{\hbar}B}~[\bar S_0,V]~ e^{-\frac{i}{\hbar}B},
\nonumber\\[2pt]
&&
C''=-t^2 H_0+2t D''+C_0+\a^2 e^{\frac{i}{\hbar}B}~V~ e^{-\frac{i}{\hbar}B}.
\eea
Thus one ultimately arrives at a free theory in which the full $\mathcal{N}=2$ Schr\"odinger superalgebra is realized in a nonlocal way.
As compared to the bosonic case, a new ingredient here is the nonlocal modification of the superconformal generators
$S$ and $\bar S$. Because our construction in section 3 is purely algebraic, it can be realized in the supersymmetric case as well.

\vspace{0.5cm}

\noindent
{\bf 5. Conclusion and outlook}\\
\noindent

In this paper we have constructed a unitary transformation which relates
a generic many--body quantum mechanics with $\mathcal{N}=2$ Schr\"odinger supersymmetry
to a set of decoupled superparticles. The simplification was achieved at a price of a highly
nontrivial and, in particular, nonlocal realization of the full
$\mathcal{N}=2$ Schr\"odinger superalgebra in a Hilbert space of a resulting free theory.
The transformation was shown to be universal and
applicable to conformal many--body models confined in a harmonic trap.
We treated explicitly the bosonic model and established a correspondence with a set of decoupled harmonic oscillators.

Let us discuss possible further developments.
As was shown above, the map allows one to draw a conclusion on the spectrum of a generic
many--body quantum mechanics with $\mathcal{N}=2$ Schr\"odinger supersymmetry.
It is interesting to study if it can provide an efficient means for constructing stationary states
of an interacting model starting from those of a decoupled theory.

In $d=2+1$ dimensions one can consider
the so called exotic Galilei algebra which involves two central charges and is intimately related to noncommutative
theories (see e.g. \cite{luk}--\cite{luk1}). It is interesting to study if the analysis of this work can
be generalized to the case of the exotic Schr\"odinger algebra. In this context see also
a recent work \cite{achp}.

Quantum integrability of many--body conformal models in higher dimensions is an exciting topic. It is tempting to
study if independent conserved charges can be extracted in a simple form from the nonlocal contribution to the generator of special
conformal transformations as prompted by our consideration in sect. 2.

Since our construction is purely algebraic,
it is likely to be applicable within the framework of
a nonrelativistic field theory. In particular, it is tempting to use the mapping in the context of
trapped Fermi gases at unitarity (see a related discussion in \cite{mehen} and references therein).

Finally, it is interesting to study a possibility of extending the present consideration to the case of
relativistic conformal algebras.

\vspace{0.5cm}

\noindent{\bf Acknowledgements}\\

\noindent
We thank P. Horv\'athy for useful comments.
This work was supported in part by RF Presidential grant
MD-2590.2008.2 and RFBR grant 08-02-90490-Ukr.

\end{document}